\documentstyle[preprint,eqsecnum,aps,epsf]{revtex}

\tightenlines
\begin{document}
\preprint{\vbox{
\hbox{INPP-UVA-99-01} 
\hbox{October, 1999} 
%\hbox{hep-ph/9910539}
}}
\draft
\def\vp{{\bf p}}
\def\ko{K^0}
\def\kb{\bar{K^0}}
\def\al{\alpha}
\def\ab{\bar{\alpha}}
\def\be{\begin{equation}}
\def\en{\end{equation}}
\def\bea{\begin{eqnarray}}
\def\eea{\end{eqnarray}}
\def\non{\nonumber}
\def\la{\langle}
\def\ra{\rangle}
\def\epp{\epsilon^{\prime}}
\def\vep{\varepsilon}
\def\to{\rightarrow}
\def\up{\uparrow}
\def\dw{\downarrow}
\def\ms{\overline{\rm MS}}
\def\ums{{\mu}_{_{\overline{\rm MS}}}}
\def\u{\mu_{\rm fact}}

\def\pr{{\sl Phys. Rev.}~}
\def\ijmp{{\sl Int. J. Mod. Phys.}~}
\def\jp{{\sl J. Phys.}~}
\def\mpl{{\sl Mod. Phys. Lett.}~}
\def\prp{{\sl Phys. Rep.}~}
\def\prl{{\sl Phys. Rev. Lett.}~}
\def\pl{{\sl Phys. Lett.}~}
\def\np{{\sl Nucl. Phys.}~}
\def\ppnp{{\sl Prog. Part. Nucl. Phys.}~}
\def\zp{{\sl Z. Phys.}~}

\title{Quark Spin and Orbital Angular Momentum in the Baryon\\}

\author{Xiaotong Song}

\address{Institute of Nuclear and Particle Physics\\ 
Department of Physics, University of Virginia, P.O.Box 400714\\
Charlottesville, Virginia 22904-4714\\}

\maketitle
\begin{abstract}
The spin and orbital angular momentum carried by different quark 
flavors in the nucleon are calculated in the SU(3) chiral quark model
with symmetry-breaking. The model is extended to all octet and decuplet
baryons. In this model, the reduction of the quark spin, due to the 
spin dilution in the chiral splitting processes, is transferred into 
the orbital motion of quarks and antiquarks. The orbital angular 
momentum for each quark flavor in the proton as function of the 
partition factor $\kappa$ and the chiral splitting probability $a$ 
is shown. Although the total amount of the quark spin reduction is 
canceled by the the equal amount increase of the quark orbital angular 
momentum, the cancellation does not apply to each quark flavor. 
Especially, the cancellation between the spin and orbital contributions 
in the baryon magnetic moment is discussed. Comparisons of our results
with other models are also shown. 

\end{abstract}

\bigskip
\bigskip
\bigskip

\pacs{12.39.Fe,~13.40.Em,~13.88.+e,~14.20.-c\\}

\newpage
%%%%%%%%%%%%%%%%%%%%%%%%%%%%%%%%%%%%%%%%%%%%%%%%%%%%%%%%%%%%%%%%%%%%%%%%
%%BEGINNING OF TEXT                           
%%%%%%%%%%%%%%%%%%%%%%%%%%%%%%%%%%%%%%%%%%%%%%%%%%%%%%%%%%%%%%%%%%%%%%%%

\leftline{\bf I. Introduction}
\smallskip

One of the important tasks in hadron physics is to reveal the
internal structure of the nucleon. This includes the study
of flavor, spin and orbital angular momentum shared by the 
quarks and gluons in the nucleon. These structures determine
the basic properties of the nucleon: spin, magnetic moment, 
axial coupling constant, elastic form factors, and the deep 
inelastic structure functions. The polarized deep-inelastic 
scattering (DIS) data \cite{emc,smc-slac-hermes,smc97} indicate 
that the quark spin only contributes about one third of the 
nucleon spin or even less. A natural and interesting question 
is: where is the {\it missing} spin ? Intuitively, and also from 
quantum chromodynamics (QCD) \cite{ji-jm}, the nucleon 
spin can be decomposed into the quark and gluon contributions
$$
{1\over 2}=<J_z>_{q+\bar q}+<J_z>_{G}={1\over 2}\Delta\Sigma+
<L_z>_{q+\bar q}+<J_z>_{G}.
\eqno (1)
$$
Without loss of generality, in (1) the proton has been chosen to 
be {\it longitudinal polarized} in the $z$ direction; it has helicity 
of $+1/2$. The angular momentum $<J_z>_{q+\bar q}$ has been
decomposed into the spin and orbital parts in (1). The total spin from 
quarks and antiquarks is $\Delta\Sigma/2=\sum\limits
(\Delta q+\Delta\bar q)/2=<s_z>_{q+\bar q}$, where 
$\Delta q\equiv q_{\up}-q_{\dw}$, and 
$\Delta{\bar q}\equiv{\bar q}_{\up}-{\bar q}_{\dw}$, and $q_{\up,\dw}$ 
(${\bar q}_{\up,\dw}$) are quark (antiquark) {\it numbers} with spin 
parallel and antiparallel to the nucleon spin, or more precisely, quark
(antiquark) numbers of {\it positive} and {\it negative} helicities.
$<L_z>_{q+\bar q}$ denotes the total orbital angular momentum carried 
by {\it quarks and antiquarks}, and $<J_z>_{G}$ is the gluon angular 
momentum. The smallness of ${1\over 2}\Delta\Sigma$ implies that the
missing part should be contributed either by the quark orbital motion 
or the gluon angular momentum. In the past decade, considerable 
experimental and theoretical progress has been made in determining 
the quark spin contribution in the nucleon \cite{review}. There is no 
direct data on $\Delta G$ except for a preliminary restriction on 
$\Delta G(x)/G(x)$ given by experiment E581/704 \cite{e581}, and 
another indirect result $\Delta G\simeq 0.5-1.5$ at $Q^2\simeq 10$
GeV$^2$ from the analysis of $Q^2$ dependence of $g_1(x,Q^2)$
\cite{smc97}. Several experiments for measuring $\Delta G$ \cite{gluon}
have been suggested. Most recently, it has been shown that 
$<J_z>_{q+\bar q}$ might be measured in the deep virtual Compton 
scattering (DVCS) \cite{ji-jws}, and one may then obtain the quark
orbital angular momentum from the difference $<J_z>-<s_z>$. Hence the 
study of the quark spin and orbital angular momentum are important and
interesting both experimentally and theoretically.

Historically, when the quark model \cite{gellmann64} was invented 
in 1960's, all three quarks in the nucleon were assumed to be in 
S-states, so $<L_z>_q=0$ and the nucleon spin arises entirely 
from the quark spin. On the other hand, in the simple parton 
model \cite{feynman69}, all quarks, antiquarks and gluons are moving 
in the same direction, i.e. parallel to the proton momentum, there is 
no transverse momentum for the partons and thus $<L_z>_{q+\bar q}=0$ 
and $<L_z>_G=0$. This picture cannot be $Q^2$ independent due to QCD 
evolution. In leading-log approximation, $\Delta\Sigma$ is $Q^2$
independent while the gluon helicity $\Delta G$ increases with $Q^2$. 
This increase should be compensated by the decrease of the orbital 
angular momentum carried by partons (see for instance Ref. \cite{bms79} 
and later analysis \cite{rat87-sd89-qrrs90}). Recently, the leading-log 
evolution of $<L_z>_{q+\bar q}$ and $<L_z>_G$, and an interesting 
asymptotic partition rule were obtained in \cite{jth96}. The `$initial$'
value of the orbital angular momenta at renormalization scale $\mu^2$ 
is determined by nonperturbative dynamics of the nucleon as a QCD bound 
state. Lattice QCD provides us with a nonperturbative tool to calculate 
the physical quantities of hadrons and has provided many interesting 
results \cite{lat1}. Meantime, many QCD inspired nucleon models 
have been developed to explain existing data and yield good physical 
insight into the nucleon. For instance, in the bag model \cite{cmbag}, 
$<s_z>_{q}\simeq 0.39$, and $<L_z>_{q}\simeq 0.11$, while in the Skyrme 
model \cite{skyrme,bali}, $\Delta G=\Delta\Sigma=0$, and $<L_z>=1/2$, 
which implies that the nucleon spin arises only from orbital motion. 

Phenomenologically, long before the EMC experimental data were published 
\cite{emc}, using the Bjorken sum rule and low energy hyperon 
$\beta$-decay data (basically axial coupling constants), Sehgal
\cite{sehgal} showed that nearly $40\%$ of the nucleon spin arises 
from the orbital motion of quarks; the remaining $60\%$ is attributed 
to the spin of quarks and antiquarks. Most recently Casu and Sehgal 
\cite{cs97} show that to fit the baryon magnetic moments and 
polarized DIS data, a large collective orbital angular momentum 
$<L_z>$, which contributes almost $80\%$ of nucleon spin, is needed. 
Hence the question of how much of the nucleon spin is coming from 
the quark orbital motion remains. This paper will discuss this 
question within the chiral quark model given in
\cite{song9705,song9802206}. In section II, the basic assumption and
formalism presented is briefly reviewed. A scheme for describing both 
spin and orbital angular momentum carried by quarks and antiquarks in 
the nucleon is discussed in section III. The important improvement is 
that the partition factor $\kappa$ is no longer restricted to be 1/3, 
which was so called `equal sharing' assumption in \cite{song9802206}. 
In this paper, the $\kappa$ can be any value in the range of [0,1/2].
Extension to the octet and decuplet baryons is given in section IV. 
The magnetic moments of the baryons are discussed in section V. 
A brief summary is given in section VI.
\bigskip

\leftline{\bf II. Chiral Quark Model}
\smallskip

The chiral quark model was first employed by Eichten, Hinchliffe and 
Quigg \cite{ehq92} to explain both the {\it sea flavor asymmetry}
$\bar d-\bar u>0$ \cite{nmc} and the {\it smallness of}
$\Delta\Sigma$ in the nucleon. The model was significantly improved 
by introducing U(1)-breaking \cite{cl1} and kaonic suppression 
\cite{smw}. A description with both SU(3) and U(1)-breakings 
was developed in \cite{song9605,wsk,cl2}. Using the low
energy hyperon decay data, the description given in \cite{song9605} 
was reformed to an one-parameter scheme in \cite{song9705} and the 
predictions are in good agreement with both spin and flavor observables. 
In this paper, we will use the notations given in \cite{song9705}.

The effective Lagrangian is 
$${\it L}_I=g_8{\bar q}\pmatrix{{G}_u^0
& {\pi}^+ & {\sqrt\epsilon}K^+\cr 
{\pi}^-& {G}_d^0
& {\sqrt\epsilon}K^0\cr
{\sqrt\epsilon}K^-& {\sqrt\epsilon}{\bar K}^0
&{G}_s^0 
\cr }q, 
\eqno (2a)$$
where ${G}_{u(d)}^0$ and ${GB}_{s}^0$ are defined as
$${G}_{u(d)}^0=+(-){{\pi^0}\over{\sqrt 2}}+
{\sqrt{\epsilon_{\eta}}}{{\eta^0}\over{\sqrt 6}}+
{\zeta'}{{\eta'^0}\over{\sqrt 3}},
~~~{G}_s^0=-{\sqrt{\epsilon_{\eta}}}{{\eta^0}\over{\sqrt 6}}+
{\zeta'}{{\eta'^0}\over{\sqrt 3}}.
\eqno (2b)$$
The breaking effects are explicitly included. $a\equiv|g_8|^2$ denotes 
the transition probability of chiral fluctuation or splitting 
$u(d)\to d(u)+\pi^{+(-)}$, and $\epsilon a$ denotes the probability 
of $u(d)\to s+K^{-(0)}$. Similar definitions are used for 
$\epsilon_\eta a$ and $\zeta'^2a$. If the breaking parameter is
dominated by mass suppression effect, one reasonably expects
$0\leq\zeta'^2a<\epsilon_{\eta}a\simeq\epsilon a\leq a$, then 
we have $0\leq\zeta'^2\leq 1$, $0\leq\epsilon_{\eta}\leq 1$, and
$0\leq\epsilon\leq 1$. We note that in our formalism, only the 
{\it integrated} quark spin and flavor contents are discussed. 

The basic {\it assumptions} of the chiral quark model are: (i) The 
quark flavor, spin and orbital contents of the nucleon are determined 
by its valence quark structure and all possible chiral fluctuations, 
and probabilities of these fluctuations depend on the interaction 
Lagrangian (2). (ii) The coupling between the quark and Goldstone 
boson is rather weak, one can treat the fluctuation $q\to q'+{\rm GB}$ 
as a small perturbation ($a\sim 0.10-0.15$) and the contributions from 
the higher order fluctuations can be neglected.  (iii) Quark spin-flip
interaction dominates the splitting process $q\to q'+{\rm GB}$. This 
can be related to the picture given by the instanton model \cite{ins}, 
hence the spin-nonflip interaction is suppressed.

Based upon the assumptions, the quark {\it flips} its spin and 
changes (or maintains) its flavor by emitting a charged (or neutral) 
Goldstone boson. The light quark sea asymmetry $\bar u<\bar d$ is 
attributed to the existing {\it flavor asymmetry} of the valence quark 
numbers (two valence $u$-quarks and one valence $d$-quark) in the 
proton. On the other hand, the quark spin reduction is due to the 
{\it spin dilution} in the chiral splitting processes. Furthermore,
the quark spin component changes one unit of angular momentum,
$(s_z)_f-(s_z)_i=+1$ or $-1$, due to spin-flip in the fluctuation 
with GB emission. The angular momentum conservation requires the 
{\it same} amount change of the orbital angular momentum but with 
{\it opposite sign}, i.e. $(L_z)_f-(L_z)_i=-1$ or $+1$. This {\it 
induced} orbital motion is distributed among the quarks and antiquarks, 
and compensates the spin reduction in the chiral splitting. This is 
the starting point to calculate the orbital angular momenta carried 
by quarks and antiquarks in the chiral quark model. We note that
the quark orbital angular momentum in the nucleon in the SU(3) 
symmetry chiral quark model was discussed in \cite{cl3}. 

For spin-up or spin-down valence $u$, $d$, and $s$ quarks, up to the 
first order fluctuation, the allowed chiral processes are
$$u_{\up,(\dw)}\to d_{\dw,(\up)}+\pi^+,~~
u_{\up,(\dw)}\to s_{\dw,(\up)}+K^+,~~
u_{\up,(\dw)}\to u_{\dw,(\up)}+{G}_u^0,~~
u_{\up,(\dw)}\to u_{\up,(\dw)}.
\eqno (3a)$$
$$d_{\up,(\dw)}\to u_{\dw,(\up)}+\pi^-,~~d_{\up,(\dw)}\to 
s_{\dw,(\up)}+K^{\rm 0},~~
d_{\up,(\dw)}\to d_{\dw,(\up)}+{G}_d^0,~~
d_{\up,(\dw)}\to d_{\up,(\dw)},
\eqno (3b)$$
$$s_{\up,(\dw)}\to u_{\dw,(\up)}+K^-,~~
s_{\up,(\dw)}\to d_{\dw,(\up)}+{\bar K}^{\rm 0},~~
s_{\up,(\dw)}\to s_{\dw,(\up)}+{G}_s^{\rm 0},~~
s_{\up,(\dw)}\to s_{\up,(\dw)}.
\eqno (3c)$$
We note that the quark helicity flips in the chiral splitting processes 
$q_{\up,(\dw)}\to q_{\dw,(\up)}$+GB, i.e. the first three processes 
in each of (3a), (3b), and (3c), but not for the last one. In the
zeroth approximation, the SU(3)$\otimes$SU(2) proton wave function
gives 
$$n^{(0)}_p(u_{\up})={5\over 3}~,~~~n^{(0)}_p(u_{\dw})={1\over 3}~,~~~
n^{(0)}_p(d_{\up})={1\over 3}~,~~~n^{(0)}_p(d_{\dw})={2\over 3}~.
\eqno (4)$$
the spin-up and spin-down quark (or antiquark) contents, up to first
order fluctuation, can be written as
$$n_p(q'_{\up,\dw}, {\rm or}\ {\bar q'}_{\up,\dw}) 
=\sum\limits_{q=u,d}\sum\limits_{h=\up,\dw}
n^{(0)}_p(q_h)P_{q_h}(q'_{\up,\dw}, {\rm or}\ {\bar q'}_{\up,\dw}),
\eqno (5)$$
where $P_{q_{\up,\dw}}(q'_{\up,\dw})$ and $P_{q_{\up,\dw}}({\bar
q}'_{\up,\dw})$ are the probabilities of finding a quark $q'_{\up,\dw}$
or an antiquark $\bar q'_{\up,\dw}$ arise from all chiral fluctuations 
of a valence quark $q_{\up,\dw}$. 
$P_{q_{\up,\dw}}(q'_{\up,\dw})$ and $P_{q_{\up,\dw}}({\bar q}'_{\up,\dw})$ 
can be obtained from the effective Lagrangian (2) and listed in Table I,
where only $P_{q_{\up}}(q'_{\up,\dw})$ and $P_{q_{\up}}({\bar
q}'_{\up,\dw})$ are shown. Those arise from $q_{\dw}$ can be obtained by
using the relations, $P_{q_{\dw}}(q'_{\up,\dw})=P_{q_{\up}}(q'_{\dw,\up})$,
$P_{q_{\dw}}({\bar q}'_{\up,\dw})=P_{q_{\dw}}({\bar q}'_{\dw,\up})$.
The notations given in Table I are defined as 
$$f\equiv{1\over 2}+{{\epsilon_{\eta}}\over 6}+{{\zeta'^2}\over 3},~~~
f_s\equiv{{2\epsilon_{\eta}}\over 3}+{{\zeta'^2}\over 3},
\eqno (6a)$$
and 
$$A\equiv 1-\zeta'+{{1-{\sqrt\epsilon_{\eta}}}\over 2},
\qquad  B\equiv \zeta'-{\sqrt\epsilon_{\eta}},\qquad
C\equiv \zeta'+2{\sqrt\epsilon_{\eta}}.
\eqno (6b)$$
The special combinations $A$, $B$, and $C$ stem from the quark and
antiquark contents in the octet and singlet neutral bosons 
${G}_{u(d)}^0$ and ${G}_{s}^0$ [see (2b)] appeared in 
the effective chiral Lagrangian (2a), while $f$ and $f_s$ stand 
for the probabilities of the chiral splittings 
$u_{\up}(d_{\up})\to u_{\dw}(d_{\dw})+{G}_{u(d)}^0$ and
$s_{\up}\to s_{\dw}+{G}_s^0$ respectively. Although there is 
no valence $s$ quark in the proton and neutron, there are one or two 
valence $s$ quarks in $\Sigma$ or $\Xi$, or other strange decuplet
baryons, and even three valence $s$ quarks in the $\Omega^-$. Hence 
for the purpose of later use we also give the probabilities arise 
from a valence $s$-quark splitting. 
In general, the suppression effects may be different for different 
baryons, hence the probabilities $P_{q_{\up,\dw}}(q'_{\up,\dw})$ 
and $P_{q_{\up,\dw}}({\bar q}'_{\up,\dw})$ may vary with the baryons.
But we will assume that they are universal for all baryons. 

Using (4), (5) and the probabilities listed in Table I, the spin-up 
and spin-down quark and antiquark contents, and the spin average 
and spin weighted quark and antiquark contents in the proton 
were obtained in \cite{song9605,song9705} and are now collected in 
Table II. For the purpose of later discussion, we write down the 
formula for the spin-weighted quark content
$$(\Delta q')^{B}=\sum\limits_q [n^{(0)}_B(q_{\up})-n^{(0)}_B(q_{\dw})]
[P_{q_{\up}}(q'_{\up})-P_{q_{\up}}(q'_{\dw})],
\eqno (7a)$$
and the spin-weighted antiquark content is zero
$$(\Delta\bar q')^B=0.
\eqno (7b)$$
Hence one has $(\Delta q)_{sea}\neq \Delta\bar q$ in the chiral quark 
model. This is different from those models, in which the sea quark and 
antiquark with the same flavor are produced as a pair from the gluon 
(see discussion in \cite{smw}). The quark spin contents in the proton are 
$$\Delta u^p={4\over 5}\Delta_3-a,~~\Delta d^p=-{1\over 5}\Delta_3-a,~~
\Delta s^p=-\epsilon a,
\eqno (7c)$$
where $\Delta_3={5\over 3}[1-a(\epsilon+2f)]$. The total quark
spin content in the proton is
$${1\over 2}\Delta\Sigma^p={1\over 2}(\Delta u^p+\Delta d^p+\Delta
s^p)={1\over 2}-a(1+\epsilon+f)\equiv {1\over 2}-a\xi_1, 
\eqno (7d)$$
where the notation $\xi_1\equiv 1+\epsilon+f$ is used. 
\bigskip

\leftline{\bf III. Quark Orbital Motion}
\smallskip

\leftline{\quad (a)~\bf Quark orbital momentum in the nucleon}
 
The quark orbital angular momentum can be discussed in a similar way. 
For instance, for a spin-up valence $u$-quark, only first three 
processes in (3a), i.e. quark fluctuations with {\rm GB} emission, 
can induce a change of the orbital angular momentum. The last process 
in (3a), $u_{\up}\to u_{\up}$ means no chiral fluctuation and it makes 
no contribution to the orbital motion and will be disregarded. The 
orbital angular momentum produced in the splitting $q_{\up}\to q'_{\dw}
+{\rm GB}$ is shared by the recoil quark ($q'$) and the Goldstone boson
(GB). If we define the fraction of the orbital angular momentum shared
by the recoil quark is $1-2\kappa$, then the orbital angular momentum 
shared by the (GB) is $2\kappa$ which, we assume, equally shared by the
quark and antiquark in the Goldstone boson. We call $\kappa$ the {\it
partition factor}, which satisfies $0<\kappa<1/2$. For $\kappa=1/3$, the
three particles, the quark and antiquark in the (GB) and the recoil quark,
equally share the induced orbital angular momentum. This is corresponding
to the `{\it equal sharing}' case discussed in \cite{song9802206}.
 
We define $<L_z>_{q'/q_{\up}}$ ($<L_z>_{{\bar q'}/q_{\up}}$) as the 
orbital angular momentum carried by the quark $q'$ (antiquark $\bar q'$), 
arises from all fluctuations of a valence spin-up quark except for no 
emission case. Considering the quark spin component changes one unit 
of angular momentum in each splitting and using Table I, we can obtain 
all $<L_z>_{q'/q_{\up}}$ and $<L_z>_{\bar q'/q_{\up}}$ for $q=u,d,s$. 
They are listed in Table III, where 
$$ \delta\equiv {{1-3\kappa}\over {\kappa}}.
\eqno (8)$$

Since the orbital angular momentum produced from a spin-up valence 
quark splitting is {\it positive}, and that from a spin-down 
valence quark splitting is {\it negative}, one has
$$ <L_z>_{q'/q_{\dw}}=-<L_z>_{q'/q_{\up}},~~ 
<L_z>_{{\bar q'}/q_{\dw}}=-<L_z>_{{\bar q'}/q_{\up}},
\eqno (9)$$
where both $q'_{\up}$ and $q'_{\dw}$ are included
in $<L_z>_{q'/q_{\up,\dw}}$ (the same is true for $<L_z>_{{\bar
q'}/q_{\up,\dw}}$).

Having obtained the orbital angular momenta carried by different 
quark flavors produced from the spin-up and spin-down valence quark
fluctuations, it is easy to write down the total orbital angular 
momentum carried by a specific quark flavor, for instance $u$-quark, 
in the proton
$$<L_z>_{u}^p=\sum\limits_{q=u,d}
[n^{(0)}_p(q_{\up})-n^{(0)}_p(q_{\dw})]<L_z>_{u/q_{\up}},
\eqno (10)$$
where $\sum$ summed over $u$ and $d$ {\it valence quarks} in
the proton. $n^{(0)}_p(q_{\up})$ and $n^{(0)}_p(q_{\dw})$ are given 
in (4). Similarly, one obtains the $<L_z>_{d}^p$, $<L_z>_{s}^p$, and 
corresponding quantities for the antiquarks. The numerical results
are listed in Table IV. Note that different baryons have different 
valence quark structure and thus different $n^{(0)}_B(q_{\up})$ and 
$n^{(0)}_B(q_{\dw})$. 

Defining $<L_z>_{q}^p$ ($<L_z>_{\bar q}^p$) as the total orbital 
angular momentum carried by {\it all quarks} ({\it all antiquarks}), 
we obtain
$$<L_z>_{q}^p\equiv<L_z>_{u+d+s}^p=(2+\delta)\kappa\xi_1a,
\eqno (11a)$$
$$<L_z>_{\bar q}^p\equiv
<L_z>_{\bar u+\bar d+\bar s}^p=\kappa\xi_1a,
\eqno (11b)$$
$$<L_z>_{q+\bar q}^p\equiv<L_z>_{q}^p+<L_z>_{\bar q}^p=\xi_1a.
\eqno (11c)$$
It means that the orbital angular momentum of each quark flavor may
depend on the partion factor $\kappa$, but the total orbital angular
momentum (11c) is independent of $\kappa$. Furthermore, the amount
$\xi_1a$ is exactly the same as the total spin reduction in (7d). 
The sum of (11c) and (7d) gives
$$<J_z>_{q+\bar q}^p=<s_z>_{q+\bar q}^p+<L_z>_{q+\bar q}^p={1\over 2}.
\eqno (11d)$$
Therefore, in the chiral fluctuations, the missing part of the quark 
spin is {\it transferred} into the orbital motion of quarks and antiquarks. 
The amount of quark spin reduction $a(1+\epsilon+f)$ in (7d) is canceled
by the equal amount increase of the quark orbital angular momentum in
(11c), and the total angular momentum of nucleon is unchanged. 

Two remarks should be made here. Although the orbital angular momentum
carried by quarks (or antiquarks) $<L_z>_{q}^p$ (or $<L_z>_{\bar q}^p$) 
depends on the chiral parameters, $a$, $\epsilon$, $\epsilon_\eta$, and
$\zeta'$, the ratio $<L_z>_{q}^p/<L_z>_{\bar q}^p=2+\delta=(1-\kappa)/
\kappa$ is {\it independent of the probabilities of chiral fluctuations}. 
For $\kappa=1/3$ (equal sharing), this ratio is 2:1. This is originated
from the mechanism of the chiral fluctuation: there are {\it two} quarks
and {\it one} antiquark in the final state. Secondly, the total {\it loss} 
of quark spin $a(1+\epsilon+f)$ appeared in Eq.(7d) is due to the fact
that there are {three} splitting processes with quark spin-flip [see the 
first three processes in (3a) and (3b)], the probabilities of these spin-flip 
splittings are $a$, $\epsilon a$, and $fa$ respectively. For the same 
reason, the total {\it gain} of the orbital angular momentum is 
$a(1+\epsilon+f)$.

The discussion can be easily extended to the neutron. Explicit 
calculation shows that $<L_z>_{u,\bar u}^n=<L_z>_{d,\bar d}^p$,
$<L_z>_{d,\bar d}^n=<L_z>_{u,\bar u}^p$, and
$<L_z>_{s,\bar s}^n=<L_z>_{s,\bar s}^p$. Using these relations, one can 
obtain the orbital angular momenta carried by quarks and antiquarks in 
the neutron. We have similar relations for $\Delta q$ from the 
isospin symmetry, hence the main results (7d), (11a-d), and related
conclusions hold for the neutron as well. Extension to other octet and 
decuplet baryons will be given in section IV.
\smallskip

\leftline{\quad (b)~\bf Numerical results}

To determine model parameters, we use similar approach given in
\cite{song9705}, where the chiral quark model with only $three$
parameters gave a good description to most existing spin and flavor 
observables. The chiral parameters $a$, $\epsilon\simeq\epsilon_\eta$, 
and $\zeta'$ are determined by three inputs, $\Delta u-\Delta d=1.26$,
$\Delta u+\Delta d-2\Delta s=0.60$, and $\bar d-\bar u=0.143$ (good 
agreement between the model prediction and spin-flavor data can be seen
from Table XIII below). The three-parameter set is: a=0.145,
$\epsilon=0.46$, and $\zeta'^2=0.10$. It gives
$$\xi_1\equiv 1+\epsilon+f=2.07.
\eqno (12)$$ 
Comparing our choice of the parameters with two extreme cases, 
$$\xi_1\equiv 1+\epsilon+f=3.0,~~~~~~~ {\rm for}~~{\rm U(3)-symmetry}~
(\epsilon=\epsilon_{\eta}=\zeta'^2=1)
\eqno (13a)$$
$$\xi_1\equiv 1+\epsilon+f=1.5,~~~~~~~ {\rm for}~~{\rm extreme~ breaking}~
(\epsilon=\epsilon_{\eta}=\zeta'^2=0)
\eqno (13b)$$
one can see that the value $\xi_1=2.07$ given in (12) is just between
those given in (13a) and (13b). 

Eq.(12) leads to
$$<L_z>_{q+\bar q}^p\simeq 0.30,
\eqno (14a)$$
and  
$$<L_z>_{q}^p=0.225,~~~~~~<L_z>_{\bar q}^p=0.075,~~~~~~({\rm for}\ \
\kappa=1/4),
\eqno (14b)$$
$$<L_z>_{q}^p=0.200,~~~~~~<L_z>_{\bar q}^p=0.100,~~~~~~({\rm for}\ \
\kappa=1/3),
\eqno (14c)$$
$$<L_z>_{q}^p=0.187,~~~~~~<L_z>_{\bar q}^p=0.113,~~~~~~({\rm for}\ \
\kappa=3/8).
\eqno (14d)$$
The orbital angular momenta shared by different quark flavors depend on
the partition factor $\kappa$ and they are listed in Table IV. We plot 
the orbital angular momenta carried by quarks and antiquarks in the proton
as function of $\kappa$ in Fig.1. Several comments should be made here. 
(1) Fig.1 shows that $<L_z>^p_s=<L_z>^p_{\bar s}$ at $\kappa=1/3$ 
and $<L_z>^p_d=<L_z>^p_{\bar d}$ at $\kappa\simeq 0.28$,
but $<L_z>^p_u$ and $<L_z>^p_{\bar u}$ cannot be equal at any value of 
$\kappa$. (2) Although the orbital angular momentum carried by each flavor
depends on $\kappa$, the total orbital angular momentum carried by the 
quarks and antiquarks does not, and is determined by the chiral
parameters (see Table IV). (3) Using the parameter set given above, 
$<L_z>_{q+\bar q}^p\simeq 0.30$, i.e., nearly $60\%$ of the proton 
spin is coming from the orbital motion of quarks and antiquarks, 
and $40\%$ is contributed by the quark and antiquark spins. Comparison 
of our result with other models is given in Table V and Fig. 2. (4) As
indicated in Eq.(11d), the total quark spin reduction (with respect to 
the SU(6) value 1/2) is canceled by the equal amount increase of the 
quark orbital angular momentum. However, the cancellation does not apply 
to each quark flavor. For example, in the NQM, $\Delta u^{p(0)}=4/3$, 
$\Delta d^{p(0)}=-1/3$, and $\Delta s^{p(0)}=0$. However, from Table IV,
we have 
$${1\over 2}\Delta u^p+<L_z>_{u+\bar u}^p=0.558~\neq~{1\over 2}\Delta
u^{p(0)}=0.667,$$
$${1\over 2}\Delta d^p+<L_z>_{d+\bar d}^p=-0.080~\neq~{1\over 2}\Delta
d^{(0)p}=-0.167,$$
$${1\over 2}\Delta s^p+<L_z>_{s+\bar s}^p=0.022~\neq~{1\over 2}\Delta
s^{(0)p}=0.$$
It is obvious that for the $u$-flavor, the orbital contribution is not 
big enough to compensate the quark spin reduction, while for the 
$d$-quark flavor, the orbital contribution is too big and the $d$-quark
spin reduction is over compensated. However, taking the sum we have
$$\sum\limits_{q=u,d,s}{1\over 2}\Delta q^p+<L_z>_{q+\bar q}^p={1\over 2}
=\sum\limits_{q=u,d,s}{1\over 2}\Delta q^{p(0)}
\eqno (11d)$$
\bigskip

\leftline{\bf IV. Extension to other Baryons.}
\smallskip

\leftline{\quad (a)~\bf Spin content in octet baryons}

We take $\Sigma^+(uus)$ as an example, other octet baryons can be 
discussed in a similar manner. The valence quark structure of 
$\Sigma^+$ is the same as the proton with the replacement
$d\to s$. Hence one has
$$n^{(0)}_{\Sigma^+}(u_{\up})={5\over 3}~,~~~
n^{(0)}_{\Sigma^+}(u_{\dw})={1\over 3}~,~~~
n^{(0)}_{\Sigma^+}(s_{\up})={1\over 3}~,~~~
n^{(0)}_{\Sigma^+}(s_{\dw})={2\over 3}~.
\eqno (15)$$
Using (5) (change $p\to \Sigma^+$), (15), and Table I, we can obtain
$\Delta u^{\Sigma^+}$, $\Delta d^{\Sigma^+}$, and $\Delta s^{\Sigma^+}$.
Similarly, we can obtain the results for $\Sigma^0$, $\Lambda^0$, and
$\Xi^0$. Those for $\Sigma^-$, and $\Xi^-$, can be obtained by using 
the isospin symmetry relations. All $\Delta q^{B}$ are listed in Table 
VI.

In general, the total spin content of quarks and antiquarks in the 
octet baryons can be written as (see Table VI)
$$<s_z>_{q+\bar q}^{B}={1\over 2}-{a\over 3}(c_1\xi_1+c_2\xi_2),
\eqno (16)$$
where $c_1$ and $c_2$ satisfy $c_1+c_2=3$, and $(c_1, c_2)$=(3, 0), 
$(4, -1)$, (0, 3), and $(-1, 4)$ for B=N, $\Sigma$, $\Lambda$, and $\Xi$
respectively. One can see that the spin reductions for all members in 
the same isospin multiplet are the same, but may be different for 
different isospin multiplets, except for the SU(3)-symmetry 
limit ($\xi_1=\xi_2=2+f$) and U(3)-symmetry limit ($\xi_1=\xi_2=3$). 
In the U(3)-symmetry limit, $<s_z>_{q+\bar q}^{N}=<s_z>_{q+\bar q}^
{\Sigma}=<s_z>_{q+\bar q}^{\Lambda}=<s_z>_{q+\bar q}^{\Xi}={1\over 2}-3a$.
Using the prameters $\xi_1\simeq 2.07$, and $\xi_2\simeq 1.27$,
we plot the quark and antiquark spin contents in different octet baryons
as function of the parameter $a$ in Fig.3. For $a\simeq 0.145$, one
obtains
$$<s_z>_{q+\bar q}^{N}\simeq 0.20,~~~
<s_z>_{q+\bar q}^{\Sigma}\simeq 0.16,~~~
<s_z>_{q+\bar q}^{\Lambda}\simeq 0.32,~~~<s_z>_{q+\bar q}^{\Xi}\simeq 0.35
\eqno (17)$$

\leftline{\quad (b)~\bf Orbital angular momentum in octet baryons}

Similar to the nucleon case, the orbital angular momenta carried 
by quarks and antiquarks in other octet baryons can be calculated.
The results for different isospin multiplets are listed in 
Table VI. The total orbital angular momentum carried by all 
quarks and antiquarks in the baryon $B$ is
$$<L_z>_{q+\bar q}^{B}={a\over 3}(c_1\xi_1+c_2\xi_2).
\eqno (18)$$
The sum of spin (16) and orbital angular momentum (18) gives
$$<J_z>_{q+\bar q}^{B}=<s_z>_{q+\bar q}^{B}+
<L_z>_{q+\bar q}^{B}={1\over 2},~~~~(B=N,~\Sigma^{\pm,0},~\Lambda^0,
\Xi^{0,-})
$$
Hence we obtain $<J_z>_{q+\bar q}^{B}=1/2$ for all octet baryons, 
i.e. the loss of the quark spin is compensated by the gain of the orbital
motion of quarks and antiquarks. The results and conclusions obtained in 
section III for the nucleon hold for other octet baryons as well. The spin
and orbital angular momentum for different quark flavors in the octet 
baryons are listed in Table VII. Using the isospin symmetry relations 
$<L_z>_{u,d}^{\Sigma^-,\Xi^-}=<L_z>_{d,u}^{\Sigma^+,\Xi^0}$ and
$<L_z>_s^{\Sigma^-,\Xi^-}=<L_z>_s^{\Sigma^+,\Xi^0}$, one can obtain
the orbital angular momenta in $\Sigma^-$ and $\Xi^-$.
Similar to the nucleon case, we have 
$<L_z>^{\Sigma^+,\Xi^0}_d=<L_z>^{\Sigma^+,\Xi^0}_{\bar d}$ at $\kappa=1/3$ 
and $<L_z>^{\Lambda}_{u,d}=<L_z>^{\Lambda}_{\bar u,\bar d}$ at
$\kappa=1/3$. This is because, for instance, the sea quark components
of $\Sigma^+$ $(uus)$ baryon are $d$ and $\bar d$, while those in the
proton are $s$ and $\bar s$. The same is true for $\Xi^0$ ($uss$) baryon.

From Table VI, one has
$$\Delta u^B-\Delta d^B=c_B[1-(\epsilon+2f)],
\eqno (19)$$
where $c_B=5/3$, $4/3$, and $-1/3$ for $B=p$, $\Sigma^+$, and
$\Xi^0$ respectively. Using the isospin symmetry relations, one 
obtains the following identity
$$\Delta u^p-\Delta u^n+\Delta u^{\Sigma^-}-\Delta u^{\Sigma^+}+
\Delta u^{\Xi^0}-\Delta u^{\Xi^-}=0.
\eqno (20a)$$
The same relation holds for $d-$quark spin and $s-$quark spin as well. 

One can show by explicit calculation that the orbital angular 
momentum $<L_z>_u^B$ in the octet baryons satisfy similar
identity
$$<L_z>_u^p-<L_z>_u^n+<L_z>_u^{\Sigma^-}-<L_z>_u^{\Sigma^+}+
<L_z>_u^{\Xi^0}-<L_z>_u^{\Xi^-}=0
\eqno (20b)$$
The same relation holds for $<L_z>_{d}^B$, $<L_z>_{s}^B$, and
$<L_z>_{\bar q}^B$. Combining (20a) and (20b), one obtains the sum rule
for the magnetic moments [see Eq.(24) below] of the octet baryons 
$$\mu_p-\mu_n+\mu_{\Sigma^-}-\mu_{\Sigma^+}+
\mu_{\Xi^0}-\mu_{\Xi^-}=0.
\eqno (21)$$
This sum rule was discussed in \cite{los} without the 
orbital contributions. Our result shows that the sum rule (21) holds 
in the symmetry breaking chiral quark model even the orbital 
contributions are included. The quark spin contents, but 
not orbital angular momentum, in the octet baryons were discussed in 
\cite{wb,los}.
\smallskip

\leftline{\quad (c)~\bf Decuplet baryons}

The above discussion can be extended to the baryon decuplet. The quark 
spin and orbital angular momenta are listed in Table VIII (we note that
the quark spin, without orbital contribution, in the decuplet baryons 
were discussed in \cite{los}). Again, the explicit calculation shows 
that $(\Delta u)^{\Delta^{-}}=(\Delta d)^{\Delta^{++}}$, 
$(\Delta u)^{\Delta^{0}}=(\Delta d)^{\Delta^{+}}$,
$(\Delta u)^{\Sigma^{*-}}=(\Delta d)^{\Sigma^{*+}}$,
and $(\Delta u)^{\Xi^{*-}}=(\Delta d)^{\Xi^{*0}}$, which are due to the
isospin symmetry of the decuplet baryon wave functions. In Table
VIII, we only list the results for $\Delta^{++}$, $\Delta^{+}$,
$\Sigma^{*+}$, $\Sigma^{*0}$, $\Xi^{*0}$, and $\Omega^{-}$. 

It is interesting to see that there is an {\it equal spacing rule} 
for total quark spin in the decuplet baryons
$$<s_z>_{q+\bar q}^{B^*}={3\over 2}-a[3\xi_1+S(\xi_1-\xi_2)],
\eqno (22a)$$
where the $S$ is the {\it strangeness} quantum number of the 
decuplet baryon $B^*$. Hence we have 
$$<s_z>_{q+\bar q}^{\Omega-\Xi}=
<s_z>_{q+\bar q}^{\Xi-\Sigma}=
<s_z>_{q+\bar q}^{\Sigma-\Delta}=a(\xi_1-\xi_2).
\eqno (22b)$$
From (22a), for the strangeless $\Delta$ multiplet, $S=0$, one obtains
$$<s_z>_{q+\bar q}^{\Delta}=3[{1\over 2}-{a\over 3}(3\xi_1)],
\eqno (22c)$$
and $<s_z>_{q+\bar q}^{\Delta}=3<s_z>_{q+\bar q}^N$, i.e. 
total spin content of $\Delta$ baryon is {\it three} times that of 
the nucleon, which is a reasonable result. We note that the equal
spacing rule (22b) was also discussed in \cite{los}.

For the orbital angular momentum (see Table VIII), we have
$$<L_z>_{q+\bar q}^{B^*}=a[3\xi_1+S(\xi_1-\xi_2)].
\eqno (23a)$$
Similar {\it equal spacing rule} holds for the orbital angular momentum
$$<L_z>_{q+\bar q}^{\Omega-\Xi}=
<L_z>_{q+\bar q}^{\Xi-\Sigma}=
<L_z>_{q+\bar q}^{\Sigma-\Delta}=-a(\xi_1-\xi_2).
\eqno (23b)$$
The sum of spin (22a) and orbital angular momentum (23a) gives
$$<J_z>_{q+\bar q}^{B^*}={3\over 2}.
\eqno (23c)$$
Once again, the spin reduction is compensated by the increase of
orbital angular momentum and keep the total angular momentum of the 
baryon (now is 3/2 for the decuplet) unchanged. 
\bigskip

\leftline{\bf V. Baryon Magnetic Moments.}

The baryon magnetic moments depend on both spin and orbital motions 
of quarks and antiquarks. In the chiral quark model all 
antiquark sea polarizations are zero, the baryon magnetic moment
can be written as
$$\mu_{B(B^*)}=\sum\limits_{q=u,d,s}\mu_q[(\Delta q)^{B(B^*)}
+<L_z>^{B(B^*)}_{q}-<L_z>^{B(B^*)}_{\bar q}]
\equiv\sum\limits_{q=u,d,s}\mu_qC_q^B ,
\eqno (24)$$
where Eq.(7b) has been used. $B(B^*)$ denote the octet (decuplet) 
baryons and $\mu_q$s are the magnetic moments of quarks. We have assumed
that the magnetic moment of the baryon is the sum of spin and orbital 
magnetic moments of individual charged particles (quarks or antiquarks). 
The assumption of {\it additivity} is commonly believed to be a good 
approximation for a loosely bound composite system, which is the basic 
description for the baryon in the effective chiral quark model. In 
addition, the baryon may contains other neutral particles, such as 
gluons (for example see discussion in \cite{bs90}). Although the gluons 
do not make any contribution to the magnetic moment, the existence of
intrinsic gluon would significantly change the valence quark structure 
of the baryon due to the spin and color couplings between the gluon and 
quarks. To calculate the baryon magnetic moments, we need to know the 
spin content $\Delta q$ (note that $\Delta\bar q=0$ in the chiral quark
model) and the difference between the orbital angular momentum
carried by quark $q$ and that carried by corresponding antiquark $\bar q$, 
which is denoted by $<L_z>^{B}_{q-\bar q}\equiv <L_z>^{B}_{q}-<L_z>^{B}_
{\bar q}$. For example, one has for $u$-quark 
$$<L_z>_{u-\bar u}^B=\sum\limits_{q}
[n^{(0)}_B(q_{\up})-n^{(0)}_B(q_{\dw})][<L_z>_{u/q_{\up}}-
<L_z>_{{\bar u}/q_{\up}}]
\eqno (25)$$
similar equation holds for $d$-quark, $s$-quark, and corresponding 
antiquarks, where $\sum$ summed over all valence quarks in the baryon 
$B$. Having obtained $\Delta q^B$ from (7a), and $<L_z>_{q-\bar q}^B$ 
from (25), one obtains the baryon magnetic moments. Since the quantities 
$P_{q_h}(q'_{\up,\dw}, {\rm or}\ {\bar q'}_{\up,\dw})$, and 
$<L_z>_{{q',{\rm or},~\bar q'}/q_{\up,\dw}}$ are known (Tables I and II)
and universal for all baryons, we only need to know the valence quark 
numbers $n^{(0)}_B(q_{\up,\dw})$ in a specific baryon $B$, and these 
numbers depend on the model of the baryon. For our purpose of showing the 
effects of the orbital angular momentum, we only consider the
valence quark structure SU(3)$_f\otimes$SU(2)$_s$ without gluon mixing.
The hybrid quark-gluon mixing model - three valence quarks and
a gluon \cite{lip} will be discussed elsewhere.
\smallskip

\leftline{\quad (a)~\bf Octet baryons}

From (7a), (24) and (25), one can obtain the analytic expressions 
of the magnetic moments for the octet baryons. It is easy to verify 
that they satisfy the following sum rules 
$$(4.70)~~~{\mu}_p-{\mu}_n={\mu}_{\Sigma^+}-{\mu}_{\Sigma^-}-
({\mu}_{\Xi^0}-{\mu}_{\Xi^-})~~~(4.22),
\eqno (26a)$$
$$(3.66)~~~-6{\mu}_{\Lambda}=-2({\mu}_p+{\mu}_n+{\mu}_{\Xi^0}+{\mu}_{\Xi^-})
+({\mu}_{\Sigma^+}+{\mu}_{\Sigma^-})~~~(3.34),
\eqno (26b)$$
$$(4.14)~~~{\mu}_p^2-{\mu}_n^2=({\mu}_{\Sigma^+}^2-{\mu}_{\Sigma^-}^2)
-({\mu}_{\Xi^0}^2-{\mu}_{\Xi^-}^2)~~~(3.56),
\eqno (26c)$$
$$(0.33)~~~{\mu}_p-{\mu}_{\Sigma^+}={3\over
5}({\mu}_{\Sigma^-}-{\mu}_{\Xi^-})
-({\mu}_n-{\mu}_{\Xi^0})~~~(0.31),
\eqno (26d)$$
where the values of the two sides taken from the data \cite{pdg96}
are shown in parentheses. The relations (26a) [or (21)] and (26b) were
first given by Franklin in \cite{frank}. The relations (26a), (26b), 
and nonlinear sum rule (26c) are not new and violated at about $10-15\%$ 
level. They have been discussed in many works, for instance \cite{karl,br,sg}.
However, the new relation (26d) is rather well satisfied. Our result 
shows that if the $SU(3)\otimes SU(2)$ valence quark structure is 
used, chiral fluctuations cannot change these sum rules even the 
orbital contributions are included. Furthermore, we have shown in 
\cite{sg} that the sum rules (26a)-(26c) also hold for more general case.  

The results of applying Eq.(24) to the magnetic moments of the baryon
octet are shown in Table IX. We find that if we choose parameters 
$\mu_u$, $\mu_d$ and $\mu_s$ by fitting to the measured values of
$\mu_p$, $\mu_n$ and $\mu_\Lambda$ as is also done in the simple SU(6) 
quark model (NQM) \cite{pdg96}, the results for all different $\kappa$
values are completely $identical$ with the NQM result. Several remarks
should be made here. (1) According to the fitting procedure, for a given
set of $\mu_{u,d,s}^{(0)}$ in the NQM, there is a corresponding set of
$\mu_{u,d,s}$ which gives the same values of $\mu_p$, $\mu_n$ and 
$\mu_\Lambda$ in the chiral quark model. (2) Seven mganetic
moments of the baryon octet satisfy four sum rules, Eqs.(26a-26d), which
hold for both the chiral quark model and the NQM, hence as soon as three
baryon magnetic moments are identical in both cases, the remaining four 
should be identical as well. (3) As we discussed in section III, the total
amount of quark spin reduction is canceled by the equal amount increase 
of the quark orbital angular momentum, $<L_z>_{q+\bar q}$, but the 
cancellation does not happen for each quark flavor. Now we turn into 
Eq.(24), the relevant term is $C_q^B\equiv\Delta q^B+<L_z>_{q-\bar
q}^B$, which differs from $\Delta q^B/2+<L_z>_{q+\bar q}^B$ in Eq.(11d),
and we have 
$$C_u^{(0)p}=4/3,\qquad C_d^{(0)p}=-1/3,\qquad C_s^{(0)p}=0,~~~~~~
({\rm NQM},~~<L_z>_{q+\bar q}=0),$$
$$C_u^{(2)p}=0.996,~C_d^{(2)p}=-0.403,~ C_s^{(2)p}=-0.067,~~
({\rm with}~<L_z>_{q+\bar q}, ~\kappa={1\over 3}),$$
$$C_u^{(1)p}=0.863,~C_d^{(1)p}=-0.397,~ C_s^{(1)p}=-0.067,~~ 
(~{\rm without}~~ <L_z>_{q+\bar q}~~).$$
The orbital contribution does move $C_u^{(1)p}$ up to $C_u^{(2)p}$, 
but it is still far below 4/3. For the $d$-quark flavor, the orbital
contribution moves $C_d^{(1)p}$ down, and $C_d^{(2)p}$ is even more 
negative than $-1/3$. Hence in the magnetic moment, Eq.(24), the
cancellation between the contributions of the quark spin reduction and 
the quark orbital angular momentum is more involved than in the total
angular momentum case [see Eq.(11d)]. 

\smallskip
 
\leftline{\quad (b)~\bf Decuplet baryons}
 
Similar to the octet baryons, the decuplet magnetic moments are 
calculated and listed in Table X. Here we use the $same$ set of 
$\mu_u$, $\mu_d$ and $\mu_s$ as in the octet sector. It is easy 
to verify that the following {\it equal spacing} rules hold,
$$\mu_{\Delta^{++}}-\mu_{\Delta^{+}}=
\mu_{\Delta^{+}}-\mu_{\Delta^{0}}=
\mu_{\Delta^{0}}-\mu_{\Delta^{-}}=
\mu_{\Sigma^{*+}}-\mu_{\Sigma^{*0}}=
\mu_{\Sigma^{*0}}-\mu_{\Sigma^{*-}}=
\mu_{\Xi^{*0}}-\mu_{\Xi^{*-}}
\eqno (27a)$$
$$\mu_{\Delta^{+}}-\mu_{\Sigma^{*+}}=
\mu_{\Delta^{0}}-\mu_{\Sigma^{*0}}=
\mu_{\Delta^{-}}-\mu_{\Sigma^{*-}}=
\mu_{\Sigma^{*0}}-\mu_{\Xi^{*0}}=
\mu_{\Sigma^{*-}}-\mu_{\Xi^{*-}}=
\mu_{\Xi^{*-}}-\mu_{\Omega^{*-}}.
\eqno (27b)$$
From Table X, the spacing is 2.82 n.m. in Eq.(27a), and $-0.36$ n.m.
in Eq.(27b). Similar to the octet baryon, the decuplet magnetic moments
with and without orbital contributions are approximately the same provided 
changing the quark magnetic moments accordingly. Hence the baryon magnetic 
moment is not a good nucleon property for revealing the quark orbital 
angular momentum, unless the quark magnetic moments are known. 

\bigskip

\leftline{\bf V. Summary}

An unified scheme for describing quark flavor, spin and orbital 
contents in the baryon in the chiral quark model is suggested. 
Contrary to the reduction effect on the quark spin component, the 
quark splitting mechanism produces a positive orbital angular 
momentum carried by the quarks and antiquarks. The results of 
the quark flavor and spin observables of the nucleon are listed 
in Table XI. They are in good agreement with the existing data. 

We have calculated the orbital angular momentum carried by different 
quark flavors in the baryons. These orbital angular momenta might be
determined indirectly in the DVCS or other processes. Attention has been
paid to the orbital contributions on the baryon magnetic moments. 

To summarize, the chiral quark model with only $a$ $few$ parameters can
well explain many nucleon properties: (1) strong flavor
asymmetry of light antiquark sea: $\bar d> \bar u$, (2) nonzero strange 
quark content, $<\bar ss>\neq 0$, (3) sum of quark spins is small, 
$<s_z>_{q+\bar q}\simeq 0.1-0.2$, (4) sea antiquarks are not polarized: 
$\Delta\bar q\simeq 0$ ($q=u,d,...$), (5) polarizations of the sea 
quarks are nonzero and negative, $\Delta q_{sea}< 0$, and (6) the 
orbital angular momentum of the sea quark is parallel to the proton 
spin. (1)-(4) are consistent with data, and (5)-(6) could be tested by 
future experiments.
\bigskip

\leftline{\bf Acknowledgments}

I would like to thank P. K. Kabir for useful comments and suggestions.
This work was supported by the Institute of Nuclear and Particle Physics,
Department of Physics, University of Virginia, and the Commonwealth of 
Virginia.
\bigskip

%{\bf References}

\bigskip

\begin{table}[ht]
\begin{center}
\caption{The probabilities $P_{q_{\up}}(q'_{\up,\dw},\bar q'_{\up,\dw})$
and $P_{q_{\up}}(q'_{\up,\dw},{\bar q}'_{\up,\dw})$} 
\begin{tabular}{cccc} 
\hline
$q'$ &$P_{u_{\up}}(q'_{\up,\dw})$ & $P_{d_{\up}}(q'_{\up,\dw})$ &
$P_{s_{\up}}(q'_{\up,\dw})$ \\ 
\hline 
$u_{\up}$ & $1-({{1+\epsilon}\over 2}+f)a+
{a\over {18}}(3-A)^2$ & ${a\over {18}}A^2$ &${a\over {18}}B^2$ \\
$u_{\dw}$ & $({{1+\epsilon}\over 2}+f)a+
{a\over {18}}(3-A)^2$ & $a+{a\over {18}}A^2$ &$\epsilon a+{a\over
{18}}B^2$ \\
$d_{\up}$ & ${a\over {18}}A^2$ &$1-({{1+\epsilon}\over 2}+f)a+
 {a\over {18}}(3-A)^2$ &${a\over {18}}B^2$ \\
$d_{\dw}$ & $a+{a\over {18}}A^2$ &
$({{1+\epsilon}\over 2}+f)a+{a\over {18}}(3-A)^2$ & 
$\epsilon a+{a\over {18}}B^2$ \\
$s_{\up}$ & ${a\over {18}}B^2$ &${a\over {18}}B^2$ &
$1-(\epsilon+f_s)a+{a\over {18}}C^2$ \\
$s_{\dw}$ & $\epsilon a+{a\over {18}}B^2$ & $\epsilon a+{a\over {18}}B^2$ 
&$(\epsilon+f_s)a+{a\over {18}}C^2$ \\
\hline
${\bar u}_{\up,\dw}$ &${a\over {18}}(3-A)^2$ & ${a\over 2}+{a\over
{18}}A^2$ &${{\epsilon a}\over 2}+{a\over {18}}B^2$ \\
${\bar d}_{\up,\dw}$ &${a\over 2}+{a\over {18}}A^2$&
${a\over {18}}(3-A)^2$ &${{\epsilon a}\over 2}+{a\over {18}}B^2$ \\
${\bar s}_{\up,\dw}$ &${{\epsilon a}\over 2}+{a\over {18}}B^2$&
${{\epsilon a}\over 2}+{a\over {18}}B^2$&
${a\over {18}}C^2$ \\
\hline
\end{tabular}
\end{center}
\end{table}

\begin{table}[ht]
\begin{center}
\caption{The spin-up, spin-down quark (antiquark), 
spin-average and spin-weighted quark (antiquark) contents in the proton.
Where $U_1={1\over 3}[A^2+2(3-A)^2]$, $D_1={1\over 3}[2A^2+(3-A)^2]$, and
$U_2=4D_2=4(\epsilon+2f-1)$.}
\begin{tabular}{ccc} 
\hline
$u_{\up}={5\over 3}+{a\over 3}(-2+{{U_1}\over 2}-{{U_2}\over 2})$& 
$d_{\up}={1\over 3}+{a\over 3}(2+{{D_1}\over 2}+{{D_2}\over 2})$& 
$s_{\up}=\epsilon a+{a\over 3}({{B^2}\over 2})$\\
$u_{\dw}={1\over 3}+{a\over 3}(5+{{U_1}\over 2}+{{U_2}\over 2})$& 
$d_{\dw}={2\over 3}+{a\over 3}(4+{{D_1}\over 2}-{{D_2}\over 2})$& 
$s_{\dw}=2\epsilon a+{a\over 3}({{B^2}\over 2})$\\
\hline 
${\bar u}_{\up}={\bar u}_{\dw}={a\over 2}+{a\over 3}({{U_1}\over 2})$&
${\bar d}_{\up}={\bar d}_{\dw}=a+{a\over 3}({{D_1}\over 2})$ &
${\bar s}_{\up}={\bar s}_{\dw}={{3\epsilon a}\over 2}+{a\over
3}({{B^2}\over 2})$\\
\hline
\hline
$u=2+{a\over 3}(3+U_1)$&$d=1+{a\over 3}(6+D_1)$& $s=3\epsilon a+{a\over
3}B^2$\\
\hline
$\bar u={a\over 3}(3+U_1)$&$\bar d={a\over 3}(6+D_1)$& $\bar s=3\epsilon
a+{a\over 3}B^2$\\
\hline
\hline
$\Delta u={4\over 3}[1-a(\epsilon+2f)]-a$ &$\Delta d={{-1}\over
3}[1-a(\epsilon+2f)]-a$ & $\Delta s=a(1-\epsilon)-a$ \\
\hline
$\Delta{\bar u}=0$ & $\Delta{\bar d}=0$ &$\Delta{\bar s}=0$ \\
\hline
%&$\Delta\Sigma=1-2a(1+\epsilon+f)$ &\\
%\hline
\end{tabular}
\end{center}
\end{table}

\begin{table}[ht]
\begin{center}
\caption{The orbital angular momentum carried by the quark $q'$ 
($\bar q'$), {\it spin-up and -down are included}, 
from a valence spin-up quark $q_{\up}$ fluctuates into all 
allowed final states.}
\begin{tabular}{cccc} 
%\hline
&$<L_z>_{q',{\bar q'}/u_{\up}}$ &$<L_z>_{q',{\bar q'}/d_{\up}}$ 
&$<L_z>_{q',{\bar q'}/s_{\up}}$\\ 
\hline 
$q'=u$ & $\kappa a[\xi_1+\delta f+{{(3-A)^2}\over 9}]$ &
$\kappa a[1+\delta+{{A^2}\over 9}]$ 
&$\kappa a[\epsilon(1+\delta)+{{B^2}\over {9}}]$ \\
$q'=d$ & $\kappa a[1+\delta+{{A^2}\over {9}}]$
&$\kappa a[\xi_1+f\delta+{{(3-A)^2}\over {9}}]$
&$\kappa a[\epsilon(1+\delta)+{{B^2}\over {9}}]$ \\
$q'=s$ &$\kappa a[\epsilon(1+\delta)+{{B^2}\over {9}}]$ 
&$\kappa a[\epsilon(1+\delta)+{{B^2}\over {9}}]$ &
$\kappa a[\xi_2+f_s\delta+{{C^2}\over {9}}]$\\
\hline
${\bar q'}={\bar u}$ &$\kappa a[{{(3-A)^2}\over {9}}]$
&$\kappa a[1+{{A^2}\over {9}}]$ 
&$\kappa a[\epsilon+{{B^2}\over {9}}]$ \\
${\bar q'}={\bar d}$ 
&$\kappa a[1+{{A^2}\over {9}}]$&$\kappa a[{{(3-A)^2}\over {9}}]$
&$\kappa a[\epsilon+{{B^2}\over {9}}]$ \\
${\bar q'}={\bar s}$&$\kappa a[\epsilon+{{B^2}\over {9}}]$ 
&$\kappa a[\epsilon+{{B^2}\over {9}}]$ 
&$\kappa a[{{C^2}\over {9}}]$ \\
\hline
\end{tabular}
\end{center}
\end{table}

\begin{table}[ht]
%\begin{table}
\begin{center}
\caption{Quark spin and orbital angular momentum in the proton in
different models.}
\begin{tabular}{ccccccc} 
%\hline
Quantity & Data \cite{smc97}   && This paper& &  Sehgal \cite{sehgal} & NQM\\ 
&& $\kappa=1/4$ & $\kappa=1/3$& $\kappa=3/8$ &&\\
\hline 
$<L_z>_u^p$       & $-$ & 0.115   & 0.130   & 0.138  & $-$    & 0  \\ 
$<L_z>_d^p$       & $-$ & 0.073   & 0.043   & 0.027  & $-$    & 0  \\ 
$<L_z>_s^p$       & $-$ & 0.038   & 0.028   & 0.023  & $-$    & 0  \\ 
$<L_z>_{\bar u}^p$& $-$ &$-$0.003 & $-$0.003&$-$0.004 & $-$    & 0  \\ 
$<L_z>_{\bar d}^p$& $-$ & 0.057   & 0.076   & 0.086  & $-$    & 0  \\ 
$<L_z>_{\bar s}^p$& $-$ & 0.021   & 0.028   & 0.031  & $-$    & 0  \\ 
\hline
$<L_z>_{q+\bar q}^p$ & $-$ & 0.30 & 0.30    & 0.30 &  0.39   & 0  \\ 
\hline
$\Delta u^p$ & $0.85\pm 0.05$ & & 0.86     &  & 0.78    & 4/3  \\ 
$\Delta d^p$ & $-0.41\pm 0.05$ & & $-0.40$ &  & $-0.34$ & $-1/3$\\ 
$\Delta s^p$ & $-0.07\pm 0.05$ & &$-0.07$ &  & $-0.14$ & 0  \\ 
\hline
${1\over 2}\Delta\Sigma^p$ & $0.19\pm 0.06$ & & 0.20 &  & 0.08 & 1/2\\ 
\hline
\end{tabular}
\end{center}
\end{table}

\begin{table}[ht]
\begin{center}
\caption{Quark spin and orbital angular momentum in different models.}
\begin{tabular}{cccccc} 
\hline 
 & NQM & MIT bag & This paper & CS \cite{cs97}&Skyrme\\ 
\hline 
$<s_z>^p_{q+\bar q}$ &1/2& 0.32 & 0.20 & 0.08&0\\
\hline 
$<L_z>^p_{q+\bar q}$ & 0 & 0.18 & 0.30 & 0.42& 1/2\\
\hline
\end{tabular}
\end{center}
\end{table}

\begin{table}[ht]
\begin{center}
\caption{The quark spin and orbital contents in the octet baryons.} 
%where $\xi_1=1+\epsilon+f$ and $\xi_2=2\epsilon+f_s$.}
\begin{tabular}{cccc} 
%\hline
 Baryon &$\Delta u^B$ & $\Delta d^B$& $\Delta s^B$ \\
\hline 
p & ${4\over 3}-{a\over 3}(8\xi_1-4\epsilon-5)$& 
$-{1\over 3}-{a\over 3}(-2\xi_1+\epsilon+5)$ & $-a\epsilon$\\
$\Sigma^+$ & ${4\over 3}-{a\over 3}(8\xi_1-5\epsilon-4)$& 
$-{a\over 3}(4-\epsilon)$ & $-{1\over 3}-{{2a}\over 3}(3\epsilon-\xi_2)$\\
$\Sigma^0$ & ${2\over 3}-{a\over 3}(4\xi_1-3\epsilon)$& 
${2\over 3}-{a\over 3}(4\xi_1-3\epsilon)$ 
& $-{1\over 3}-{{2a}\over 3}(3\epsilon-\xi_2)$\\
$\Lambda^0$ & $-a\epsilon$&  $-a\epsilon$& 
$1-2a(\xi_2-\epsilon)$\\
$\Xi^0$ & $-{1\over 3}-{a\over 3}(-2\xi_1+5\epsilon+1)$& 
$-{a\over 3}(4\epsilon-1)$ & ${4\over 3}-{a\over 3}(8\xi_2-9\epsilon)$\\
\hline
   &$<L_z>_q^B$ & $<L_z>_{\bar q}^B$&$<L_z>_{q+\bar q}^B$\\
\hline 
p & $(2+\delta)\kappa a\xi_1$& $\kappa a\xi_1$&$a\xi_1$
\\
$\Sigma^+$ & $(2+\delta){{\kappa a}\over 3}(4\xi_1-\xi_2)$& 
${{\kappa a}\over 3}(4\xi_1-\xi_2)$& ${a\over 3}(4\xi_1-\xi_2)$
\\
$\Lambda^0$&$(2+\delta)\kappa a\xi_2$& $\kappa a\xi_2$&$a\xi_2$
\\
$\Xi^0$ & $(2+\delta){{\kappa a}\over 3}(4\xi_2-\xi_1)$& 
${{\kappa a}\over 3}(4\xi_2-\xi_1)$& ${a\over 3}(4\xi_2-\xi_1)$
\\
\hline
\end{tabular}
\end{center}
\end{table}

\begin{table}[ht]
\begin{center}
\caption{Quark spin and orbital angular momentum in other octet baryons.}
\begin{tabular}{cccccccccc} 
%\hline
 Baryon   &    & $\Sigma^+$ &  &   & $\Lambda$ & & &$\Xi^0$& \\
&$\kappa=1/4$ & $\kappa=1/3$ &$\kappa=3/8$ &$\kappa=1/4$ &$\kappa=1/3$ 
& $\kappa=3/8$ & $\kappa=1/4$ & $\kappa=1/3$ &$\kappa=3/8$\\
\hline 
$<L_z>_u^B$  &0.130&0.141&0.147&0.038&0.028&0.023&0.014&0.000&$-$0.008\\ 
$<L_z>_d^B$  &0.096&0.071&0.058&0.038&0.028&0.023&0.023&0.017&0.014\\
$<L_z>_s^B$  &0.029&0.015&0.007&0.063&0.067&0.069&0.071&0.080&0.085\\
\hline 
$<L_z>_{\bar u}^B$&0.005&0.007&0.008&0.021&0.028&0.031&0.025&0.033&0.037\\
$<L_z>_{\bar d}^B$&0.053&0.071&0.079&0.021&0.028&0.031&0.013&0.017&0.019\\
$<L_z>_{\bar s}^B$&0.026&0.035&0.039&0.004&0.006&0.007&$-$0.001&$-$0.001& 
$-$0.002\\ 
\hline
$<L_z>_{q+\bar q}^B$ & 0.34&0.34&0.34 &0.18&0.18&0.18 &0.15& 0.15&0.15  \\ 
\hline
$\Delta u^B$ & &0.84    & & & $-$0.07 & & & $-$0.29 &    \\ 
$\Delta d^B$ & &$-$0.17 & & & $-$0.07 & & & $-$0.04 &    \\ 
$\Delta s^B$ & &$-$0.35 & & &    0.77 & & &   1.05  &  \\ 
\hline
${1\over 2}\Delta\Sigma^B$ & & 0.16& & & 0.32& &  & 0.35&  \\ 
\hline
\end{tabular}
\end{center}
\end{table}

\begin{table}[ht]
\begin{center}
\caption{The quark spin and orbital contents in the decuplet baryons.}
\begin{tabular}{cccc} 
 Baryon &$\Delta u^{B^*}$ & $\Delta d^{B^*}$& $\Delta s^{B^*}$
%&$\Delta\Sigma^{B^*}$ 
\\
\hline 
$\Delta^{++}$ & $3-3a(2\xi_1-\epsilon-1)$& $-3a$ & $-3a\epsilon$
%&$3-2a(3\xi_1)$
\\
$\Delta^{+}$ & $2-a(4\xi_1-2\epsilon-1)$& $1-a(2\xi_1-\epsilon+1)$
&$-3a\epsilon$
%& $3-2a(3\xi_1)$
\\
$\Sigma^{*0}$ & $1-2a\xi_1$& $1-2a\xi_1$&$1-2a\xi_2$
%& $3-2a(2\xi_1+\xi_2)$
\\
$\Sigma^{*+}$ & $2-a(4\xi_1-\epsilon-2)$& $-a(\epsilon+2)$ &$1-2a\xi_2$
%& $3-2a(2\xi_1+\xi_2)$
\\
$\Xi^{*0}$ & $1-a(2\xi_1+\epsilon-1)$& $-a(2\epsilon+1)$ 
&$2-a(4\xi_2-3\epsilon)$
%& $3-2a(\xi_1+2\xi_2)$
\\
$\Omega^{-}$ & $-3a\epsilon$& $-3a\epsilon$ &
$3-6a(\xi_2-\epsilon)$
%& $3-2a(3\xi_2)$
\\
\hline
&$<L_z>_q^{B^*}$ & $<L_z>_{\bar q}^{B^*}$& $<L_z>_{q+\bar
q}^{B^*}$
%& $<J_z>_{q+\bar q}^{B^*}$ 
\\
\hline 
$\Delta$&${{(2+\delta)\kappa a}}(3\xi_1)$&${{\kappa a}}(3\xi_1)$&$a(3\xi_1)$
%&${3\over 2}-a(1-3k)(3\xi_1)$
\\
$\Sigma$&${{(2+\delta)\kappa
a}}(2\xi_1+\xi_2)$&${{\kappa a}}(2\xi_1+\xi_2)$&$a(2\xi_1+\xi_2)$
%& ${3\over 2}-a(1-3k)(2\xi_1+\xi_2)$
\\
$\Xi$& ${{(2+\delta)\kappa a}}(\xi_1+2\xi_2)$& ${{\kappa a}}(\xi_1+2\xi_2)$
&$a(\xi_1+2\xi_2)$
%& ${3\over 2}-a(1-3k)(\xi_1+2\xi_2)$
\\
$\Omega$&${{(2+\delta)\kappa a}}(3\xi_2)$& 
${{\kappa a}}(3\xi_2)$ &$a(3\xi_2)$
%& ${3\over 2}-a(1-3k)(3\xi_2)$
\\
\hline
\end{tabular}
\end{center}
\end{table}

\begin{table}[ht]
\begin{center}
\caption{Comparison of our predictions with data for the octet baryon
magnetic moments. The naive quark model (NQM) results are also listed. 
The quantity used as input is indicated by a star.} 
\begin{tabular}{cccccc} 
 Baryon    & data & & This paper  &  & NQM \\
     &           &$\kappa=1/4$&$\kappa$=1/3& $\kappa=3/8$& \\
\hline 
p     &     $ 2.79\pm 0.00$&  & 2.79$^{*}$ & &2.79$^{*}$\\
n     &     $-1.91\pm 0.00$&  &$-1.91^{*}$ & &$-1.91^{*}$\\
$\Sigma^+$ &$ 2.46\pm 0.01$&  & 2.67       & &  2.67\\
$\Sigma^-$ &$-1.16\pm 0.03$&  &$-1.09$     & &$-1.09$ \\
$\Lambda^0$&$-0.61\pm 0.00$&  &$-0.61^{*}$ & &$-0.61^{*}$ \\ 
$\Xi^0$    &$-1.25\pm 0.01$&  & $-1.43$    & &$-1.43$\\
$\Xi^-$    &$-0.65\pm 0.00$&  & $-0.49$    & &$-0.49$\\
\hline
$\mu_u$&  &2.404    &2.351    &2.328   &1.85 \\
$\mu_d$&  &$-1.047$ &$-0.944$ &$-0.892$&$-0.97$ \\
$\mu_s$&  &$-0.657$ &$-0.623$ &$-0.606$&$-0.61$ \\
\hline
\end{tabular}
\end{center}
\end{table}

\begin{table}[ht]
\begin{center}
\caption{Comparison of our predictions with data for the decuplet baryon 
magnetic moments. The naive quark model (NQM) results are also listed.} 
\begin{tabular}{cccc} 
 Baryon & data    & This paper  &  NQM \\
  &   & $\kappa=1/3$  &   \\
\hline 
$\Delta^{++}$ &$4.52\pm 0.50\pm 0.45$ \cite{bos91}&5.55&5.58 \\
 & $3.7~<~\mu_{\Delta^{++}}~<~7.5$ \cite{pdg96} &     &\\
$\Delta^{+}$  & $-$                   &  2.73  &  2.79 \\
$\Delta^{0}$  & $-$                   &$-0.09$ &  0.00 \\
$\Delta^{-}$  & $-$                   &$-2.91$ &$-2.79$\\ 
$\Sigma^{*+}$ & $-$                   &  3.09  &  3.11 \\
$\Sigma^{*0}$ & $-$                   &  0.27  &  0.32 \\
$\Sigma^{*-}$ & $-$                   &$-2.55$ &$-2.47$\\
$\Xi^{*0}$    & $-$                   &  0.63  &  0.64 \\
$\Xi^{*-}$    & $-$                   &$-2.19$ &$-2.15$\\
$\Omega^{-}$  &$-1.94\pm 0.17\pm 0.14$ \cite{diehl91}&$-1.83$&$-1.83$\\
 &$-2.024\pm 0.056$ \cite{wal95}& & \\
 &$-2.02\pm 0.05$ \cite{pdg96}  & & \\
\hline
$\mu_u$&  &2.351      &1.85 \\
$\mu_d$&  &$-0.944$   &$-0.97$ \\
$\mu_s$&  &$-0.623$   &$-0.61$ \\
\hline
\end{tabular}
\end{center}
\end{table}

\begin{table}[ht]
\begin{center}
\caption{Quark spin and flavor observables in the proton. The quantity
used as input is indicated by a star. }
\begin{tabular}{cccc} 
\hline
Quantity & Data& This paper &NQM\\
\hline 
$\bar d-\bar u$ & $0.147\pm 0.039$ \cite{nmc} & $0.143^*$  &0\\
& $0.100\pm 0.018$ \cite{e866} & &  \\
${{\bar u}/{\bar d}}$ & $[{{\bar u(x)}\over {\bar d(x)}}]_{x=0.18}=0.51\pm
0.06$ \cite{na51}& 0.64 &   $-$\\
 & $[{{\bar u(x)}\over {\bar d(x)}}]_{0.1<x<0.2}=0.67\pm 0.06$ \cite{e866}
& &\\
${{2\bar s}/{(\bar u+\bar d)}}$ & ${{<2x\bar s(x)>}\over {<x(\bar
u(x)+\bar d(x))>}}=0.477\pm 0.051$ \cite{ccfr95}& 0.72 & $-$\\
 ${{2\bar s}/{(u+d)}}$ & ${{<2x\bar s(x)>}\over
{<x(u(x)+d(x))>}}=0.099\pm0.009$ \cite{ccfr95} & 0.13 &0\\
 ${{\sum\bar q}/{\sum q}}$ & ${{\sum<x\bar q(x)>}\over{\sum<xq(x)>}}
=0.245\pm 0.005$ \cite{ccfr95} & 0.23 &0\\
 $f_s$ & $0.10\pm 0.06$ \cite{gls91} & 0.10 & 0\\
       & $0.15\pm 0.03$ \cite{dll95} &      &    \\
       & ${{<2x\bar s(x)>}\over {\sum<x(q(x)+\bar q(x))>}}
=0.076\pm 0.022$ \cite{ccfr95} & &  \\
$f_3/f_8$ & $0.21\pm 0.05$ \cite{cl1} & 0.22 & 1/3\\
\hline 
$\Delta u$ & $0.85\pm 0.05$ \cite{smc97} & 0.86 & 4/3\\
$\Delta d$&$-0.41\pm 0.05$ \cite{smc97} &$-$0.40&$-$1/3\\
$\Delta s$&$-0.07\pm 0.05$ \cite{smc97} &$-0.07$&0\\
$\Delta\bar u$, $\Delta\bar d$ & $-0.02\pm 0.11$ \cite{smc96} &0&0 \\
$\Delta_3/\Delta_8$ &2.17$\pm 0.10$&2.12& 5/3\\
$\Delta_3$ &1.2601$\pm 0.0028$ \cite{pdg96}&1.26$^{*}$& 5/3\\
$\Delta_8$& 0.579$\pm 0.025$ \cite{pdg96}& 0.60$^{*}$&1 \\
\hline
\end{tabular}
\end{center}
\end{table}

\begin{figure}[h]
\epsfxsize=5.0in
\centerline{\epsfbox{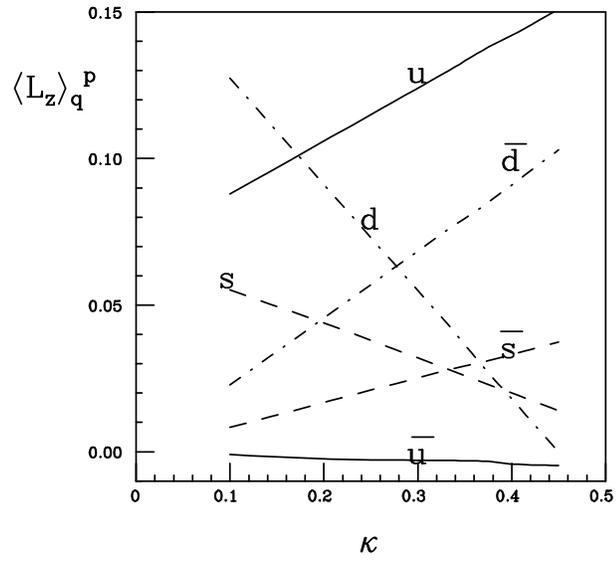}}
\caption{Quark or antiquark orbital angular momentum $<L_z>_{q,\bar q}$
in the proton as function of the partition factor $\kappa$.}
\end{figure}

\begin{figure}[h]
\epsfxsize=5.0in
\centerline{\epsfbox{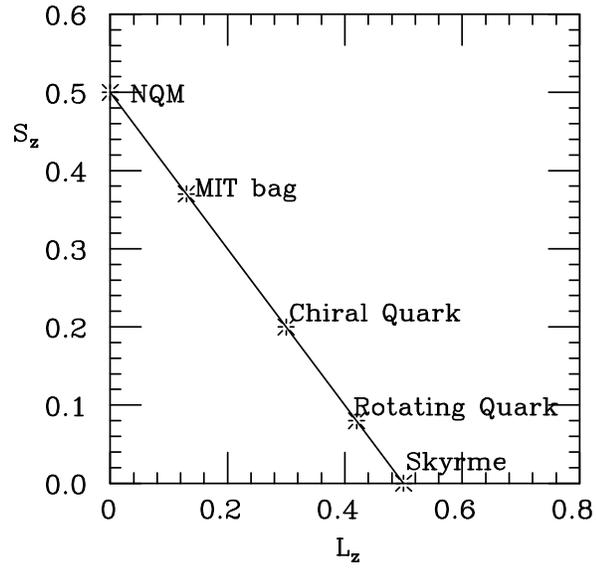}}
\caption{Quark spin and orbital angular momentum ($<s_z>_{q+\bar q}$
versus $<L_z>_{q+\bar q}$) in the nucleon in different models.}
\end{figure}

\begin{figure}[h]
\epsfxsize=5.0in
\centerline{\epsfbox{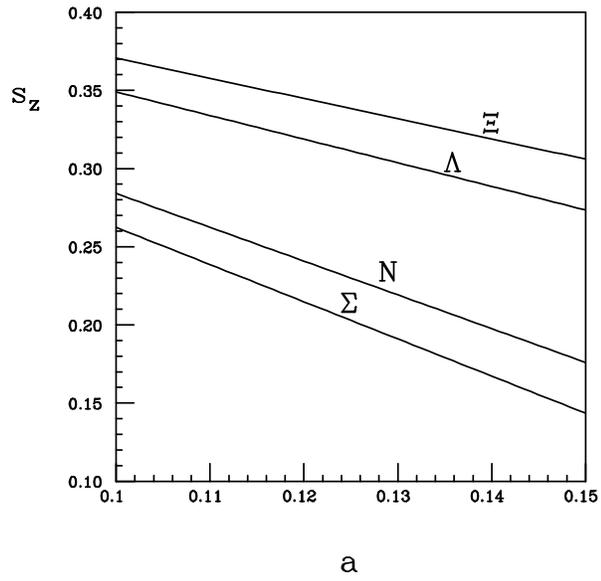}}
\caption{Quark spin content ($<s_z>_{q+\bar q}^B$) in different octet
baryons as function of $a$}
\end{figure}

\end{document}